\newcommand{\q}{\vec q}
\newcommand{\x}{\vec x}
\newcommand{\xii}{{\vec x}_I}
\newcommand{\xs}{{\vec x}_S}

\documentclass[prl,aps]{revtex4}
\usepackage{graphics}
\begin{document}
\title{Entangled imaging and wave-particle duality: from the microscopic to the macroscopic realm}
\author{A.~Gatti, E.~Brambilla and L.~A.~Lugiato}
\affiliation{INFM, Dipartimento di Scienze CC.FF.MM.,
Universit\`a dell'Insubria, Via Valleggio 11, 22100 Como, Italy}
\begin{abstract}
We formulate a theory for entangled imaging, which includes
also the case of a large number of photons in the two entangled beams.
We show that the results for imaging
and for the wave-particle duality features, which have been demonstrated
in the microscopic case, persist
in the macroscopic domain.
We show that the  quantum character
of the imaging phenomena is guaranteed by the simultaneous
spatial entanglement in the near and in the far field.
\end{abstract}
\centerline{Version \today}
\maketitle
A major trend in physics at present is to ``push the realm
of quantum physics well into the macroscopic world" \cite{bib1}.
In the case of photon systems, quantum effects
such as squeezing or twin beams can be found also in the regime of large
photon number. However, the analysis of fundamental
phenomena, such as wave-particle duality, for a macroscopic
electromagnetic field requires a spatially multi-mode
treatment, as developed in \cite{bib3,bib4}.\\
The field of quantum imaging \cite{bib5} and, especially, the
topic of entangled two-photon imaging (EPI) provides an
ideal framework for such a discussion.
The theory of EPI was pioneered by Klyshko \cite{bib6} who formulated
a heuristic approach, that stimulated a number of key experiments,
especially in the laboratory of Shih \cite{bib7}. Similar experiments
in the group of Zeilinger \cite{bib1} addressed the discussion of fundamental
issues in quantum physics.
Recently, the Boston group formulated a systematic theory of such phenomena \cite{bib8}.
All the papers \cite{bib6,bib7,bib8} concerned the regime of single photon-pair
detection in the parametric downconversion process. In this work
we formulate a theory for EPI which holds for arbitrary downconversion efficiency, and therefore
encompasses also the case in which a large number of photons are detected in each pump
pulse, as it happens e.g. in the experiment described in \cite{bib9}.
A key role in our analysis is played by the concept of spatial entanglement
\cite{bib10,Brambilla} we introduced previously.
Special attention is devoted to a clear identification of the features
that require the presence of quantum entanglement rather than classical correlations,
as those of the experiment of Boyd et al.\cite{Boyd}.\\ \indent
We consider a type II $\chi^{(2)}$ crystal of length $l_c$. We assume 
an undepleted pump beam, with a Gaussian profile of waist $w_p$, and a Gaussian temporal pulse profile
of duration $\tau_p$. 
In the parametric down-conversion process the
signal/idler ($S/I$) photon pairs are emitted over a broad band
of temporal frequencies (bandwidth $\Omega_0\propto\l_c^{-1}$) and a
broad band of spatial frequencies (bandwidth $q_0\propto(\lambda l_c)^{-1/2}$, with $\lambda$ being
the central wavelength of the down-converted fields).
We developed a numerical model, based on the Wigner representation,
that simulates the propagation  of the three waves inside a realistic $\chi^{(2)}$ 
crystal,
including the effects of 
diffraction, spatial and temporal walk-off, and temporal dispersion \cite{bib13}.
Here we will not describe such a method, but we will focus on
some key results for a  conceptual
experimental scheme suited to discuss the wave-particle aspects at a macroscopic level.
We also used an analytic approach, valid in the limit of a plane-wave cw pump,  
where propagation inside the crystal is described in terms of the 
unitary transformation  \cite{bib3,bib13}:
\begin{eqnarray}
a_i^{out} (\q, \Omega)&=& U_i(\q,\Omega) a_i^{in}(\q,\Omega) + V_i(\q,\Omega) a_j^{in \: \dagger} (-\q, -\Omega)
\nonumber \\ & & \qquad i\ne j=S,I \; 
						\label{inout}
\end{eqnarray}
linking $S/I$ fields at the input with those at the output face of the nonlinear crystal. 
$a_i(\q, \Omega)^{in/out}$ are annihilation operators of plane-wave modes,
$\q$ being the transverse wave-vector and  $\Omega$ the  shift from the carrier frequency.
The explicit form of the functions 
 $U_i$ and $V_i$ is given for example in \cite{bib3}.
For brevity, we drop in the following the frequency argument from all the 
formulas (even if we took it into account in our calculations).
When the transformation equivalent to (\ref{inout}) in a Schroedinger-like  picture  
is applied to the input vacuum state
of the $S/I$ fields, we obtain the output entangled state:
\begin{eqnarray}
\left| \psi \right\rangle
 = \prod_{\q}
\left\{
\sum_{n=0}^{\infty} c_n(\q)
|n,\q \rangle_S
|n,-\q \rangle_I
\right\} 
							\label{thestate}\;,
\end{eqnarray}
where 
$|n,\q\rangle_{S/I}$ denotes a 
Fock state with $n$ photons in mode $\q$ of the $S/I$ beam, and 
$c_n(\q)=\left\{ U_S(\q)V_I(-\q)\right\}^n  \left| U_S(\q)\right|^{-(2n+1)}$. Moreover, 
\begin{equation}
\left|c_n(\q)\right|^2=\frac{\langle n (\q) \rangle^n}{\left[1+\langle n (\q) \rangle\right]^{n+1}}\;,
\label{eq2}
\end{equation}
where $\langle n (\q)\rangle $ is the average number of photons in
mode $\q$. Almost all down-conversion literature  is limited
to the case $\langle n (\q) \rangle \ll 1$, in which state~(\ref{thestate})
reduces to
\begin{eqnarray}
|\psi \rangle%_{out} 
\approx  \prod_{\q}c_0 (\q)
|0,\q\rangle_S
|0,-\q\rangle_I 
+ \sum_{\q}  \left\{   c_1(\q) |1,\q\rangle_S |1,-q\rangle_I 
\prod_{\q_1 \ne \q}c_0(\q_1)
|0,\q_1\rangle_S
|0,-\q_1\rangle_I \right\}
								\label{eq3}
\end{eqnarray}
In this case, which we  refer to as  the {\em microscopic} case, one
detects coincidences of single photon pairs; in application to imaging,
the image is reconstructed from a statistics over a large number
of coincidences.
In this paper we focus, instead, on the case in which the average photon number per mode
is not negligible,
so that all the terms of the expansion (\ref{thestate}) are relevant 
(we call it the {\em macroscopic} case). In this case the entanglement is with respect to photon number, and this model
predicts ideally perfect correlations in $S/I$ photon number detected in two symmetric
modes $\q$ and $-\q$ \cite{Brambilla}.\\
An interesting analytical limit is that of a short crystal
(where diffraction and walk-off of the $S/I$ fields along the crystal become negligible), where
the output state can be written in the form
\begin{equation}
\label{eq4}
|\psi\rangle%_{out}
=\prod_{\x}
\left\{
\sum_{n=0}^{\infty}
c_n(\q=0)|n,\x\rangle_S
            |n,\x\rangle_I
\right\}\; ;
\end{equation}
$\x$ denotes position in the transverse plane at the crystal exit (``near-field"), and 
$|n,\x\rangle$ is the Fock state with $n$ photons at point $\x$. In this limit one has
ideally a perfect correlation in the number of $S/I$ photons detected at the same near field position.
We incidentally note that if only one beam of the two is considered, 
its reduced density matrix 
%obtained by tracing out the degrees of freedom of the other beam from 
%the density matrices corresponding to the state (\ref{thestate}), or(\ref{eq4})
is diagonal in the Fock state basis, and  corresponds 
to a thermal statistics  with average number of photons given by (\ref{eq2}).\\
In the more sophisticated numerical model, the finite size
of the pump has the effect that, if
one idler photon is emitted in direction $\q$, its twin photon will travel in
the symmetric direction $-\q$, within an uncertainty $\delta q \approx 1/w_p$, which hence represents
the uncertainty in the signal transverse momentum when determined from a measurement
of the idler transverse momentum.
On the other hand, due to the finite length  of the crystal, twin photons created in a single
down-conversion process at the same position, are separated by diffraction 
along the crystal. Hence the uncertainty in the position of
a signal photon conditioned to the detection of an idler  photon at position $\x$ 
is given by a coherence length $l_{coh}=1/q_0 \approx (\lambda \l_c/2\pi)^{\frac{1}{2}}$.
Really important for the purpose of imaging is
%is the number of coherence areas
%contained inside the beam transverse cross-section, which gives an estimate of 
the number of pixels that can be resolved in an imaging scheme based on correlation measurements.
This number is assessed, both in the near and in the far field, by the ratio $(w_p/l_{coh})^2=(q_0/\delta q)^2$.
The simultaneous presence of entanglement both in momentum and in position 
is a fundamental property of the down-converted photons, which, as we will see,
plays a crucial role in the imaging process. This property persists for a
large photon number, a case in which the entanglement assures a 
photon number spatial correlation at the quantum level both in the near and the far field.\\
\begin{figure}
\scalebox{0.6}{
\includegraphics{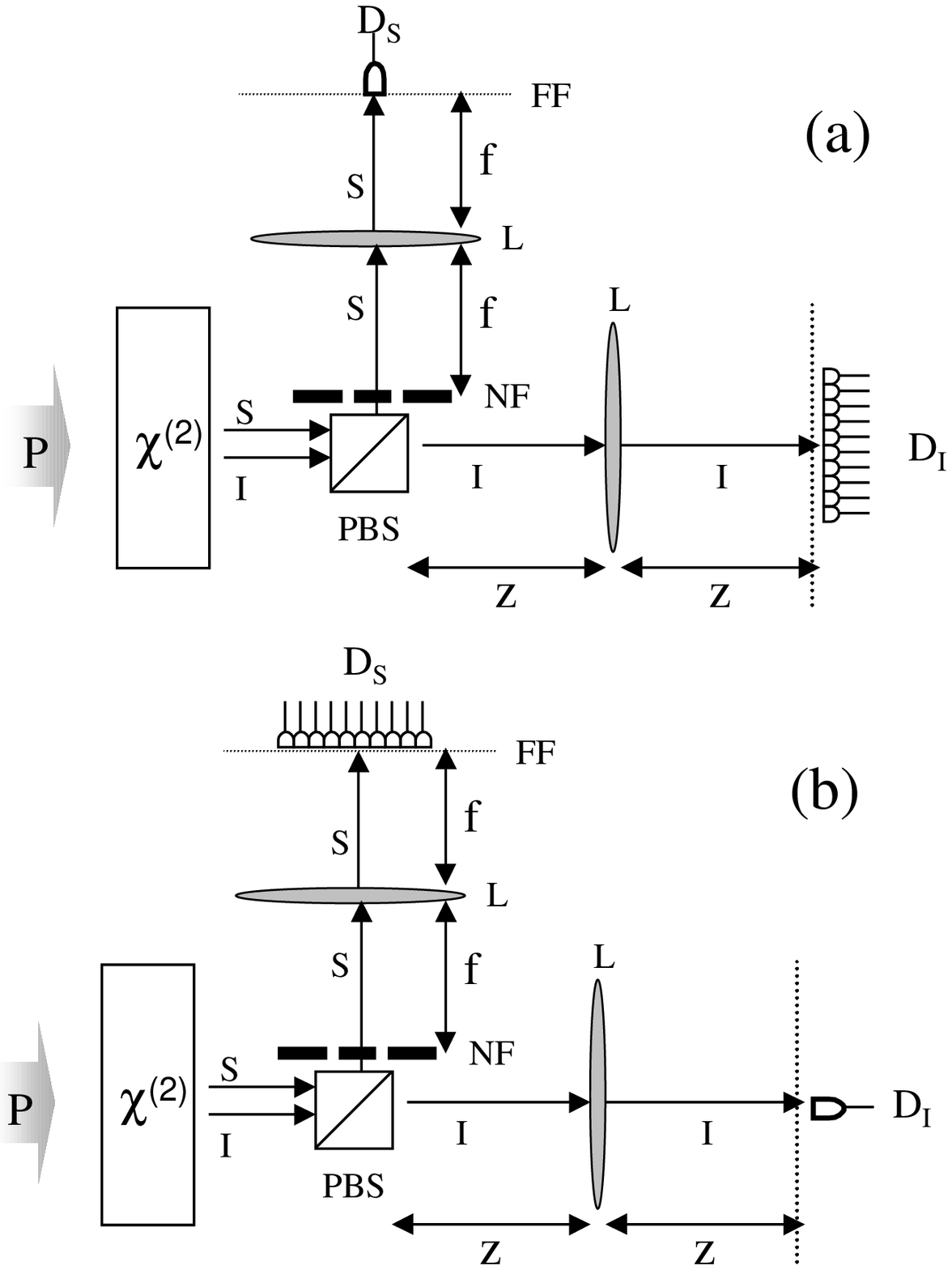}}
\caption{(a)Imaging scheme. $P=$ pump beam, $\chi^{(2)}=$ type II crystal, $S=$
signal field, $I=$ idler field, $D_S, D_I=$  detectors, $L=$ lens
with focal length $f$, $NF=$ near field, $FF=$ far field, $PBS=$
polarizing beam splitter. (b) scheme for the discussion of  
fundamental aspects.}
\label{fig2}
\end{figure}
Figure \ref{fig2} illustrates a compact imaging scheme. The $S/I$ beams are separated
by the polarizing beam splitter PBS (we assume for simplicity that the distance 
from the crystal exit  to PBS is negligible). 
In the path of the $S$ beam there is an object, which is
imaged by a lens on the far-field plane, where it is detected by a single  
point-like detector $D_S$. 
An identical lens images the  $I$ beam on its detection plane, where it is 
observed by an array of detectors $D_I$. The distance $z$ between the PBS
and  the lens, and between the lens and $D_I$ can be varied; we focus on the cases $z=f$ and
$z=2f$, in which we  will see that the diffraction pattern ($z=f$)  and the image ($z=2f$) 
of the object can be reconstructed
by correlation measurements.
For definiteness, we discuss the case in which the object is a double slit,
with $a$ being the width of the two slits and $d$ their distance. For $z=f$,
Fig.\ref{fig2}~a corresponds to the setup of some of the experiments in \cite{bib7}.\\
The object is in the path of the $S$ beam, which is observed by a 
point-like detector, 
and no information about it can be
obtained by direct detection.  
As a straightforward  generalization from the coincidence measurements 
of the microscopic case\cite{bib6,bib7,bib8}, we consider the spatial correlation 
of the $S/I$ detected intensities. Precisely,  we denote with
$I_S(\xs)$ and $I_I(\xii)$ the intensities detected by $D_S$ and by the array $D_I$
averaged over a detection time $\tau_D$ (in typical pulsed experiments $\tau_D\approx \tau_P$),
and we introduce the spatial correlation function
\begin{equation}
\langle I_I(\xii)I_S(\xs)\rangle=  \langle I_I(\xii)\rangle \langle I_S(\xs)\rangle
+
\langle \delta I_I(\xii)\,\delta I_S(\xs)\rangle \, .
					\label{G2}
\end{equation}
The object information is contained 
in the correlation of intensity fluctuations 
$G(\xii, \xs)=\langle \delta I_I(\xii) \delta I_S(\xs)\rangle$
as a function of $\xii$ for fixed $\xs$. 
Toghether with the scheme (a)
of Fig.\ref{fig2}, we consider the alternative scheme (b)
in which, conversely, the $S$ beam is detected by  an array,  and $I$
by a point-like detector. Such a scheme was
analysed in \cite{bib1} in the microscopic case.
In the macroscopic case the image is provided by $G(\xii, \xs)$
as a function of $\x_S$ for fixed $\xii$.\\
Let us first consider the case  $z=f$, in scheme (b).
If  the $I$ field is not detected, there is no possibility of 
observing the interference fringes by direct measurement of field S alone,
unless the object is  contained in a coherence area, i.e. $d\ll l_{coh}$.
In the microscopic case it was argued
\cite{bib1} that in principle one could detect the $I$ photon, and obtain ''which-path"
information on the $S$ photon, and this is enough to cancel the fringes. More in general,
we  argue that, since the $S$ beam alone  is in a incoherent thermal mixture, 
the interference fringes are not visible due to the lack
of coherence.
However, in order to make fringes visible,  it is enough to condition the $S$ beam  measurement 
 to a measurement
of the $I$ beam  by a single  point-like detector.  
In the microscopic case the fringes are observed via coincidence measurements, 
as explained in \cite{bib1}, because detection
of the $I$ photon in the far field determines  
the $S$ photon momentum before the double slit, due to momentum entanglement, 
providing a quantum erasure \cite{bib14} of any which-path
information. In the general case, the basic mechanism 
is the $S/I$ far field intensity correlation, and 
calculations performed with the analytical model (\ref{inout}) show that
\begin{eqnarray}
G(\xs,\xii) &\propto&
\left|\tilde{T} \left(\q=\frac{2\pi}{\lambda f}(\xs+\xii)\right) \right|^2 \,
\left|U_S \left(\frac{-2\pi\xii }{\lambda f} \right)V_I \left(\frac{2\pi\xii}{\lambda f} 
		\right)\right|^2 
\:,							\label{eq6}
\end{eqnarray}
where $\tilde{T}(\q)$ is the Fourier transform of the transmission function $T(\x)$ describing
the object. Under the conditions $a \gtrsim l_{coh}$, $d<w_p$, the entire interference-diffraction
pattern is visible with good resolution. The result (\ref{eq6})
is symmetric with respect to $\xs$ and $\xii$,
 hence the same pattern appears in the imaging scheme (a).
\begin{figure}
\scalebox{0.4}{
\includegraphics{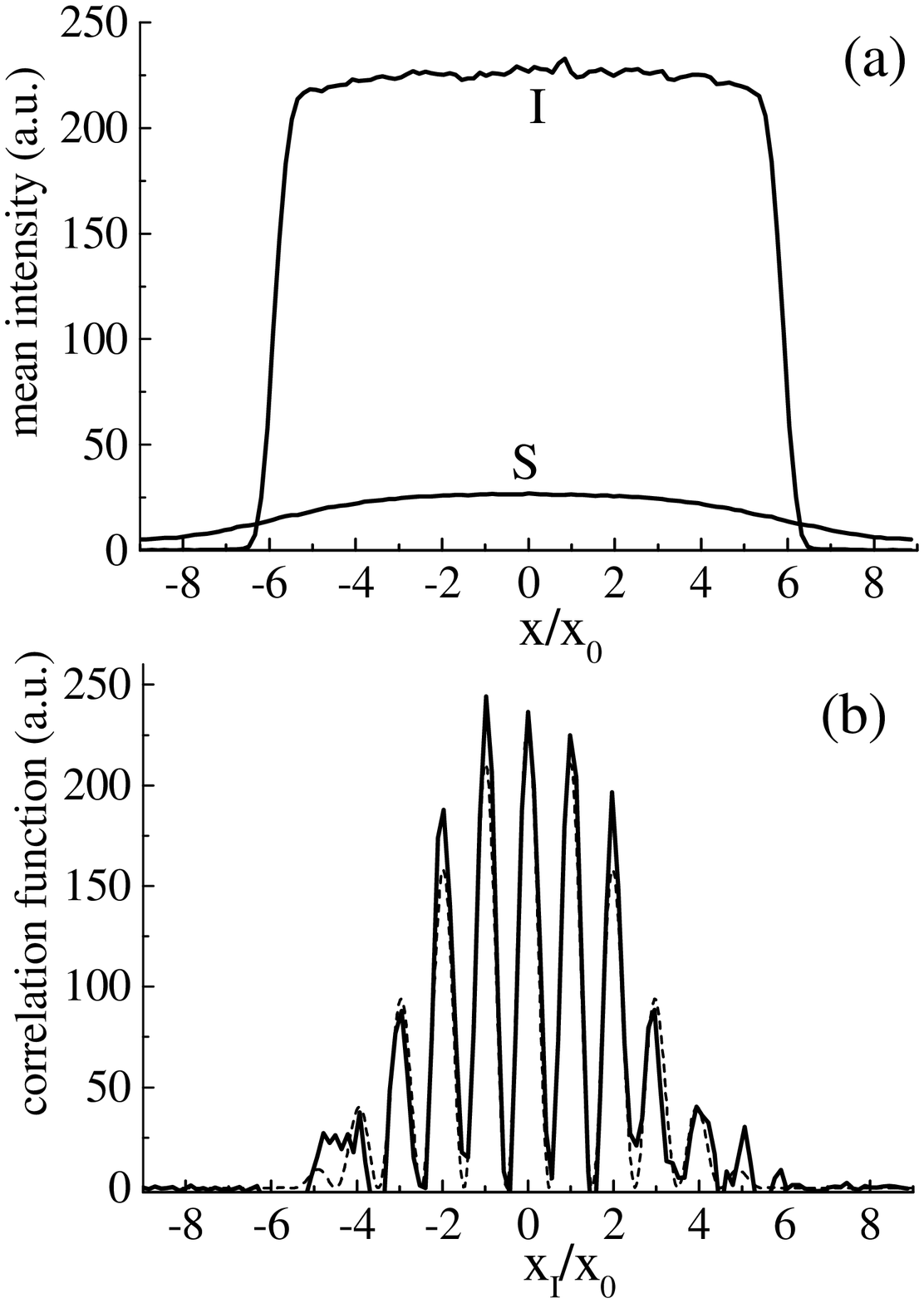}}
\caption{Numerical simulation of the experiment in Fig.~2a with $z=f$.
 Parameters are those of a 4 mm BB0 crystal, with $w_p=332\mu$m, 
$\tau_p=1.5$ps ($\tau_{coh}=0.87$ps, $l_{coh}=16.6$ $\mu$m), $a=17\mu$m, $d=104\mu$m. 
(a)Mean intensity
of the $S/I$ beams after 10000 pulses; (b) Solid line: correlation 
 $G(x_I,x_S)$
as a function of $x_I$ after 10000 pulses; dashed line: plane-wave result of Eq.(\ref{eq6}).
$x_0= \lambda f q_0/(2\pi)$}
\label{fig3}
\end{figure}
Fig.\ref{fig3}b shows the result of a 1-D numerical simulation of the pattern reconstruction via
intensity correlation function. 
A statistical average over 
a reasonable number of pump shots was enough, because $\tau_p$ was on the same order of magnitude 
as the amplifier coherence time  $\tau_{coh}=1/\Omega_0$.
%\footnote{Here the coherence time is
%%defined as $\tau_{coh}= \frac{l_c}{v_g^S}- \frac{l_c}{v_g^I}$, where $v_g^S$,$v_g^I$ are the group velocities
%of the two waves inside the crystal}
\begin{figure}[h]
\scalebox{0.25}{
\includegraphics{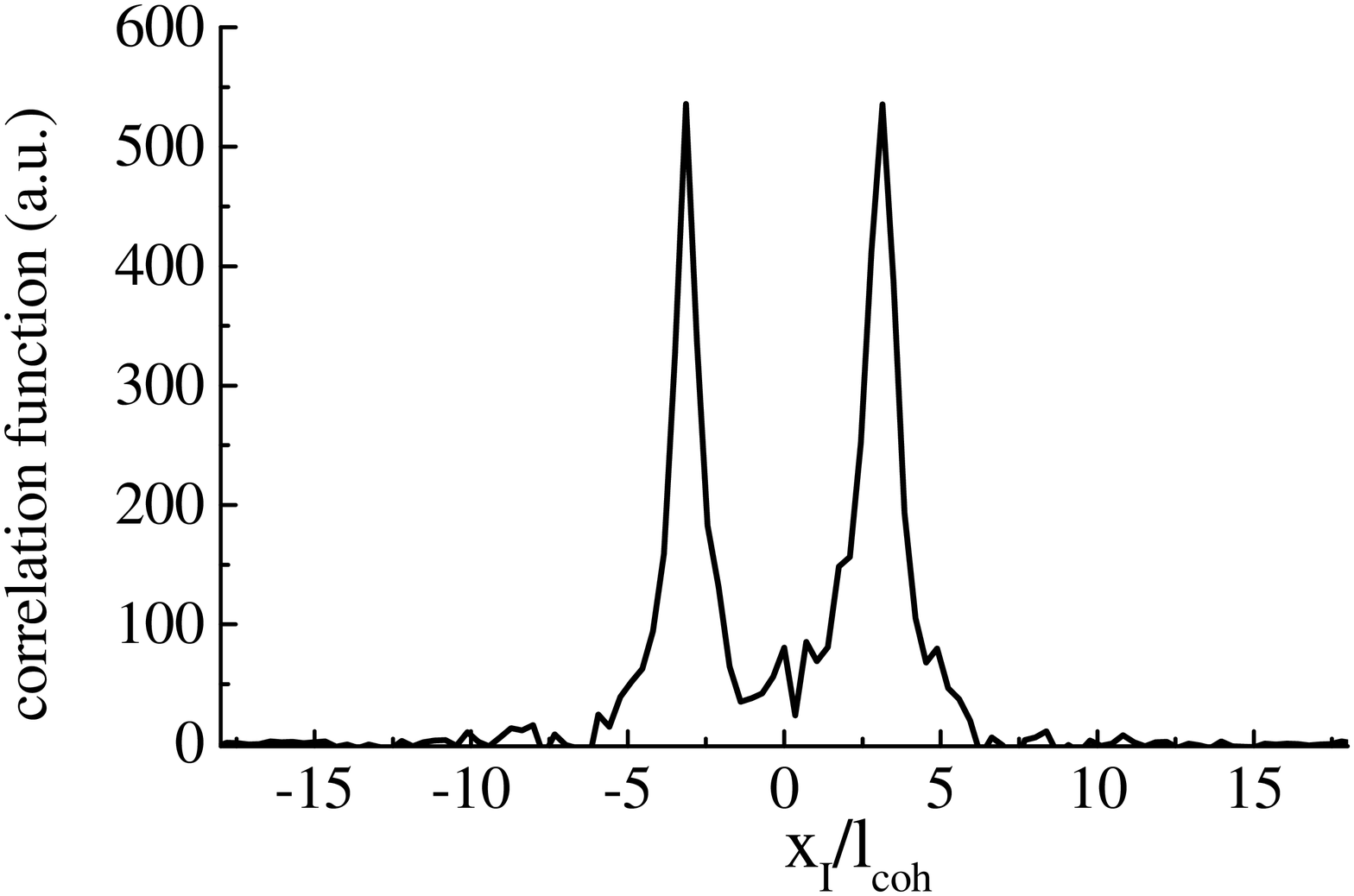}}
\caption{Numerical simulation of the experiment in Fig.~2a with $z=2f$.
Parameters as in figure \ref{fig3} The figure shows $G(x_I, x_S)$
as a function of $x_I$ after 10000 pulses}
\label{fig4}
\end{figure}
Our calculations show that when $\tau_D\gg \tau_{coh}$,
the first term at rhs of  Eq. (\ref{G2}) becomes much larger
than the second term, that contains all the information about the object, at the expenses 
of the visibility.\\
Consider now scheme (b) in the $z=2f$  case, in which 
the $D_I$ detector lies in the image plane with respect to the object, and 
the measurement exploits the $S/I$ spatial correlation in the near-field.
In the microscopic case fringes are not visible because the detection of the $I$ photon in
the near field, due to position entanglement, 
provides perfect ''which path" information about the $S$ photon \cite{bib1}.
Our general result is that, again, for $a > l_{coh}$, $d<w_p$ 
\begin{equation}
\label{eq7}
G(\xs,\xii) 
\propto
|T(\xii)|^2 \left|U_S \left(\frac{2\pi\xs }{\lambda f} \right)V_I \left(-\frac{2\pi\xs}{\lambda f} 
		\right)\right|^2 .
\end{equation}
In scheme (b), where $\xii$ is fixed, there is no information about the object.
However, in scheme (a) ($\xs$ fixed)the object image can be reconstructed via 
the correlation measurement. Hence, by only changing
the optical setup in the path of the idler, {\em which does not go through the object}, 
one is able to pass from the diffraction pattern to the image of an object.
This result is confirmed by our numerical simulation shown in Fig.\ref{fig4}.\\
In order to assess the quantum nature of the phenomena observed in the imaging scheme (a),
the key question is whether these results can be reproduced by using a ``classical" mixture,
instead of the pure entangled state (\ref{thestate},\ref{eq4}).
It is natural to focus on the two mixtures
\begin{eqnarray}
& W=\prod_{\q}
\left\{
\sum_{n=0}^{\infty}
|c_n(\q)|^2
|n,\q\rangle_S|n,-\q\rangle_{II}
\langle n,\q|_S\langle n,-\q|
\right\}				\label{W}\\
& W'=\prod_{\x}
\left\{
\sum_{n=0}^{\infty}
|c_n(0)|^2
|n,\x\rangle_S|n,\x\rangle_{II}
\langle n,\x|_S\langle n,\x|
\right\}
				\label{Wprime}
\end{eqnarray}
Mixture (\ref{W}) preserves the local $S/I$ spatial intensity correlation
in the far field, while the intensity correlation function is completely 
delocalized in the near field. By following the same notation 
of \cite{bib8}, we indicate by $h_j(\x_j, \x)$ the linear Kernel
describing propagation throught the imaging setup of beam $j=I,S$; we 
introduce their Fourier transforms 
%$  \tilde{h}_j (\x_j, \q ) = \int \frac{d \x}{2\pi} exp{(i \q \cdot \x)}h_i(\x_j, \x)$, 
 $  \tilde{h}_j (\x_j, \q )$
describing how a $\q$ component of the $j$ beam at the crystal exit face is transformed
into the field at point $\x_j$ at the detection plane. With mixture (\ref{W}) we obtain
\begin{equation}
G(\xs, \xii) =\int {d \q} |\tilde{h}_S(\xs, \q)|^2 |\tilde{h}_I(\xii, -\q)|^2 
| U_S(\q)V_I(-\q) |^2. 
	\label{GW}
\end{equation}
The optical setup in the $S$ beam arm is fixed, and
\begin{equation}
\tilde{h}_S (\xs, \q) =  \frac{1}{i\lambda f} \tilde{T}\left( \frac{2\pi}{\lambda f}\xs - \q \right) 
	\label{GW2}
\end{equation}
In the $z=f$ configuration of Fig.2a, $|\tilde{h}_I (\xii, -\q)|^2 \propto \delta 
(\frac{2\pi}{\lambda f}\xii + \q )$ and we obtain the same result of Eq.(\ref{eq6}). 
Hence  
fringes are visibile with the classical mixture (\ref{W}) in the same way  as with 
the pure EPR state (\ref{thestate}).  
However, for $z=2f$,  $ \tilde{h}_I (\xii, -\q) \propto \exp{(i \xii \cdot \q)}$, 
and the correlation function is constant with $\xii$; thus in this case 
the scheme gives no information at all about the object. Conversely,
the mixture (\ref{Wprime}) preserves the $S/I$ local
intensity correlation only in the near field. Not surprisingly,
 in this case  the $z=2f$  scheme (a) provides  the image of the object,
as with the pure state, 
but in the $z=f$ case the fringes are not visibile. The key point is that only the
pure EPR state (\ref{thestate},\ref{eq4}) displays  $S/I$ spatial correlation 
both in the near and in the far field.
This analysis is not in contrast with the basic conclusion of Ref.\cite{Boyd}, 
that the result of each single experiment
in EPI can be reproduced by a classically correlated source.
 Here we argue that only in the presence of quantum entanglement 
the whole set of results illustrated in Fig.2b and 3 can be obtained {\em by using a single source},
and {\em  by keeping the optical setup in the signal beam arm fixed.}\\ 
In conclusion, we formulated a theory that encompasses both the microscopic
(single photon pair detection) and the macroscopic (multi-photon detection) case.
Our results 
show that the imaging and wave-particle duality phenomena, 
observed in the microscopic case, persist
in the macroscopic domain, and indicate a possible experiment that is able to discriminate
between the presence of quantum entanglement or classical correlation in the two beams.
 Clearly, there is a pratical limit in the macroscopic level
that can be  attained preserving such phenomena. In order to increase
the number of down-converted photons, usually either the pump beam is more focused 
($w_p$ is decreased),
or the crystal length $l_c$ is increased. However, when the condition 
$w_p = l_{coh} \propto \sqrt{\lambda l_c}$ is reached, the resolution 
of the spatial entanglement in the near and
far field, as well as in the entangled imaging, is  completely lost.
\section*{Acknowledgements}
Work in the framework of the FET project QUANTIM of the EU.

\end{document}